\documentclass[aps,prd,amsmath,showpacs,nofootinbib]{revtex4}
\RequirePackage[]{amssymb}
\usepackage{graphicx}

\def\({\left(}
\def\){\right)}

\newcommand{\beq}{\begin{equation}}
\newcommand{\eeq}{\end{equation}}
\newcommand{\bea}{\begin{eqnarray}}
\newcommand{\eea}{\end{eqnarray}}

\newcommand{\bean}{\begin{eqnarray*}}
\newcommand{\eean}{\end{eqnarray*}}
\newcommand{\bs}{\begin{subequations}}
\newcommand{\es}{\end{subequations}}

\newtheorem{opr}{Definition}

\newcommand{\diag}{\textrm{diag}}
\DeclareMathOperator{\const}{const}

\newcommand{\artanh}{\textrm{artanh}}

\begin{document}


\title{Static cylindrically symmetric dyonic wormholes \\
in 
6-dimensional  Kaluza--Klein theory: Exact solutions}

\author{Asya V. Aminova}
\email{asya.aminova@kpfu.ru}%
\affiliation{Department of Relativity Theory and Gravity, Kazan Federal University,
18 Kremlyovskaya St.,
Kazan 420008,
Russian Federation
}
\author{Pavel I. Chumarov}%
\email{gpagvva@gmail.com}%
\affiliation{Department of Relativity Theory and Gravity, 
Kazan Federal University,
18 Kremlyovskaya St.,
Kazan 420008,
Russian Federation
} %

\begin{abstract}

 We study cylindrically symmetric Abelian  wormholes (WhC) in 
$(4+n)$-dimensional Kaluza--Klein theory.  It is shown that static, four-dimensional, cylindrically symmetric solutions in  $(4+n)$-dimensional  Kaluza--Klein theory with maximal Abelian isometry group $U(1)^n$ of the internal space 
with diagonal internal metric
can be obtained, as in the case of a supersymmetric static black hole \cite{Cv}, only if the isometry group of the internal space is broken down to the $U (1)_e \times U(1)_m$ gauge group; they correspond to dyonic configurations  with one electric $(Q_e)$ and one magnetic $(Q_m)$ charge  
 that are related either  
to 
the same $U(1)_e$ or $U(1)_m$  gauge field  or to  
different factors of the $U (1)_e \times U(1)_m$  gauge group of the effective 
 6-dimensional Kaluza--Klein theory.
We find new exact solutions of the 
 6-dimensional Kaluza--Klein theory
with two Abelian gauge fields,  dilaton and scalar fields, associated with the internal metric.
 We  obtain new types of cylindrically symmetric wormholes supported by
the radial and longitudinal electric and magnetic fields. 

\end{abstract}


 \maketitle

\section{Introduction}

As a starting point we recall the comment 
in a paper \cite{CvKK} where a class of static spherically symmetric solutions in $(4+n)$-dimensional Kaluza--Klein theory with Abelian isometry was studied: "
We assumed that the internal isometry group $G$ is Abelian. In this case, different supersymmetric static spherical solutions spontaneously break $G$ down to different $U(1)_E \times U(1)_M$ factors as the vacuum configurations. We suspect that the same thing  will happen for axially symmetric stationary configurations, but it remains to be proven". 
This article is motivated by the desire to find out to what extent the numerous results (e.g. \cite{Gib}--\cite{5}, and others) obtained for (spherically symmetric) black holes in 
higher-dimensional unified theories,  related   
to supergravity and string theory, can be transferred to  
axially symmetric wormholes. We started with a cylindrically symmetric wormhole; 
the first task was
derivation    from the higher-dimensional theory of an effective 4-dimensional theory of pure gravity, including  
``external gravitons" -- the space-time metric,  and 
``internal gravitons" -- the scalar and gauge fields associated with the extra dimensions, as well as  finding the exact solutions of the resulting theory.
This problem is solved in this paper. The study of  geodesic structure,  singularity structure and thermal properties of the wormhole solutions we found, as well as  inclusion of fermions in the theory and finding  connections between Kaluza-Klein wormhole solutions and string theory, would be the next step in the research of cylindrically symmetric wormholes within supersymmetric unified field theories.

By wormholes one usually means 
topological features in the form of
``handles''  (throats) connecting 
different regions of the space--time continuum.
 
The typically-discussed wormhole models   are endowed with spherical
symmetry (see, for example, the
survey~\cite{Visser}).
It has been shown~\cite{3} that spherically symmetric wormholes can exist only
in the presence of  
so-called ``exotic matter'', which refers to a variety of  field
configurations 
that have, for example,
negative energy density and negative pressure.

A cylindrically symmetric space-time has a preferred direction -- the
axis of (axial)  symmetry. Examples of axially symmetric systems include  
cosmic strings, among others.
We consider a static cylindrically symmetric space-time metric~\cite{1}
\begin{equation}\label{cm}
ds ^ 2 = e ^ {2 \gamma (u)} dt ^ 2 - e ^ {2 \omega (u)} du ^ 2 - e ^ {2 \xi (u)}
dz ^ 2 - e ^ {2 \beta (u)} d \phi ^ 2,
\end{equation}
where 
$ u $ is an arbitrary cylindrical
radial  coordinate, $ z \in (- \infty, + \infty) $ is the longitudinal
coordinate, and
  $ \phi \in [0, 2 \pi] $ is the angular coordinate. The ``circle  radius''  $
R(u): = e ^ {\beta (u)} $ is non-negative and tends to $ + \infty $ when $ u
\rightarrow \pm \infty $.
Space-time~\eqref{cm} has  
one timelike Killing vector $ {\xi} _1 =
\partial_ t $ and  two spacelike Killing vectors $  {\xi} _2 =
\partial_ \phi $,
$  {\xi} _3 = \partial_z $, which define the axial symmetry.
By a suitable choice of coordinate $ u $ the equality 
$$
\omega (u) = \beta (u) +\gamma (u) +\xi (u)
$$
can be satisfied
(in what follows we assume that this equality holds), and  from \eqref {cm} we
have
\begin{equation}\label{cm0}
ds ^ 2 = e ^ {2 \gamma (u)} dt ^ 2 - e ^ {2[\beta (u) +\gamma (u) +\xi (u)]  }
du ^ 2 - e ^ {2 \xi (u)} dz ^ 2 - e ^ {2 \beta (u)} d \phi ^ 2.
\end{equation}

Following K.  Bronnikov and J.  Lemos ~\cite{1} we 
make the definitions:
\begin{opr}\label{opr1}
We say that the metric~\eqref{cm} describes a cylindrically symmetric wormhole
if the circle radius  $R(u)$
has an absolute minimum  $R(u_0) > 0$ at some point $u = u_0$ and for all
possible values of $u$ the metric functions $\omega(u)$, $\beta(u)$,
$\gamma(u)$, $\xi(u)$ in~\eqref{cm} are smooth and finite. 
\end{opr}
\begin{opr}\label{opr2}
The throat of a cylindrically symmetric wormhole with metric~\eqref{cm} is a
cylindrical
hypersurface defined by the equation
$$
u = u_0.
$$
\end{opr}

It has been shown in ~\cite{1}  that the existence of the static, cylindrically symmetric wormholes does not require violation of the weak or null  energy conditions near the throat, and the cylindrically symmetric geometry of  wormhole configurations
can be generated by less exotic sources compared to 
the case of spherical symmetry.
The exact   solutions of Einstein's theory of gravity with scalar, spinor, and
electromagnetic fields
describing  
static cylindrically symmetric wormholes with metric~\eqref{cm0}
 have been obtained in~\cite{1}, and  all the solutions are not
 asymptotically flat. 
It has been proved that  in the absence of material
fields that violate the weak or null energy
conditions, i. e. in the case of everywhere nonnegative energy density of matter,  flat asymptotic behavior on both sides of a 
 cylindrically symmetric
wormhole is impossible~\cite{1}.

In this article we discuss static, cylindrically symmetric space-times \eqref{cm0}  within 
$(4+n)$-dimensional Kaluza--Klein theory with Abelian isometry,  and our results confirm the validity of the above  ``no--go" statement.

The paper is organized as follows. In  section II 
 we discuss dimensional reduction of  
$(4 + n)$-dimensional  gravity with a $4$-dimensional space-time metric \eqref{cm0} and 
 gauge and scalar fields that are compatible with cylindrical symmetry. We show that  the isometry group $U (1)^n$  of the internal 
space with  diagonal  metric is broken down to the $U (1)_e\times 
 U (1)_m$ 
 gauge group and obtain  the constraints on charges in the case of compactification on a $2$-torus. 
In section III we derive and integrate equations of motion for cylindrically symmetric configurations with two Abelian gauge fields, dilaton   and  scalar fields. We find the 
 exact solutions describing cylindrically symmetric
wormholes in the
six-dimensional effective Kaluza--Klein theory with radial electric and magnetic fields.
In section IV the 
exact solutions describing
cylindrically symmetric 6-d Kaluza--Klein  wormholes with the longitudinal electric and magnetic
fields are found.
Conclusions are given in Section V.

In this paper we use units in which the speed of the light  in  
vacuum $ c$  and
gravitational constant $ G  $ are taken 
to be one: $c=G=1$.

\section{DIMENSIONAL REDUCTION OF (4 + N)-D GRAVITY WITH DIAGONAL INTERNAL METRIC}

The effective 4-dimensional Kaluza-Klein  theory is obtained from $(4 +n)$-dimensional pure gravity with the Einstein-Hilbert action
~\cite{Cho}:
\begin{equation}\label{dejstvie}
S_{4+n}= \frac {1}{16 \pi G_ {4 + n}} \int \sqrt {-g ^ {(4 + n)}} R ^ {(4 + n)}
d ^ {4 + n} x,
\end{equation}
by compactifying the $n$ extra spatial coordinates on a compact manifold.
Here $ G_ {4 + n} $ is the gravitational constant in $ 4 + n $ dimensions, $ R^{ (4+n)}$ is the Ricci scalar and $g^ {(4+n)}$ is the determinant  of a  metric $g_{AB}$  \footnote{Upper-case letters $A, B,\ldots$ are used for the indices of the coordinates $ x ^ 0, \ldots, x ^ {3 + n} $ in a $
(4 + n) $-dimensional space. The  Greek letters $ \mu, \nu, \ldots $
 denote the indices of coordinates $ x ^ 0 $, $ x ^ 1 $, $ x ^ 2 $, $ x
^ 3 $ of the four-dimensional space-time and the lower-case letters $a, b,\ldots$  are used for the coordinates
$ x ^ 4, \ldots, x ^ {3 + n} $ of an internal space.
}
of the unified space $M^{4+n}$.

The dimensional reduction of the (4+n)-d gravity is achieved by requiring 
right invariance of
the metric $g_{AB}$
under the action of an isometry group $G_n$ with Killing vector fields $X_a\  (a = 1 ,2, ... , n)$ and structure constants $f^c_{ab}$:
$L_{X_a}g_{AB}=0,\quad   [X_a,X_b] =f^c_{ab}X_c.
$
 $M^{4+n}$ is considered as  a principal fiber bundle $P(M^4,G_n)$ with the four-dimensional space-time $M^4$ as the base manifold and $G_n$  as the structure group. In a local direct-product coordinate basis ($\partial_\mu, \partial_a$)   the metric $g_{AB}$
 is written as
$$
 \begin{pmatrix}
e ^ {- \sigma / \alpha} g_ {\mu \nu} + e ^ { 2\sigma/(n\alpha)  } \rho_ {bc} A_ \mu ^ b A_ \nu ^ c & \ e ^ {2 \sigma /(n \alpha)}  \rho_ {cb} A_ \mu ^ c \\ \\
e ^ {2 \sigma / (n\alpha)} \rho_ {ac} A_ \mu ^ c & \ e ^ {2 \sigma
/ (n\alpha)}  \rho_ {ab}
\end{pmatrix},
$$
where   $\sigma$ is dilaton field and $A^1_\mu,\ldots, A^n_\mu $ are  gauge potentials,
$\alpha = \sqrt{1+2/n}$ is
 the coupling constant of the dilaton  to the gauge fields, $\rho_{ab}$ is the unimodular part  of the internal metric: $\det(\rho_{ab})=1$.
The  dependence of $g_{AB}$ on internal space coordinates is determined by its right invariance under the action of  $G_n$ \cite{CvKK}, \cite{Cho}:
$$\partial_ag_{\mu\nu}=0,\quad  \partial_aA^c_\mu=- f^c_{ab}A^b,\quad
\partial_a\rho_{bc}= f^d_{ab}\rho_{dc}+  f^d_{ac}\rho_{bd}.
$$
When the isometry group $G_n$ is unimodular, the $(4+n)$-d Lagrangian density of the Einstein-Hilbert action (\ref{dejstvie}) becomes  explicitly independent of internal coordinates. The dimensional reduction is obtained by integrating over the internal space and, up to a total divergence, the Lagrangian
of the effective four-dimensional Kaluza-Klein  theory is given by \cite{Cho}
\begin{equation}\label{hyper-lagr}
L=\frac{1}{16\pi}\sqrt{-g}[R+e^{-\alpha\sigma}\tilde R-\frac{1}{4}e^{\alpha\sigma}\rho_{ab}F^a_{\mu\nu}F^{b\ \mu\nu}-\frac{1}{2}\partial_\mu \sigma \partial^\mu \sigma-\frac{1}{4}\rho^{ab}\rho^{cd}D_\mu\rho_{ac}D^\mu\rho_{bd}+\lambda(\det(\rho_{ab})-1)],
\end{equation}
where $g=\det(g_{\mu\nu})$,  $\tilde R$ is the Ricci scalar of the unimodular part $\rho_{ab}$ of the internal metric, $F^a_{\mu\nu}=\partial_\mu A^a_\nu- \partial_\nu A^a_\mu -\hat g f^a_{bc}A^b_\mu A^c_\nu$ is the field strength of  $A^a_\mu$, $\hat g$ is the gauge coupling constant, $D_\mu\rho_{ab} = \partial_\mu \rho_{ab} - f^d_{cb}A^c_\mu \rho_{ad}$ is the corresponding gauge covariant derivative,  $\rho^{ac}\rho_{cb}=\delta^a_b$, and $\lambda$  is a Lagrange multiplier; the 4-d gravitational constant  has been set equal to 1. When $G_n$ is 
  Abelian isometry group (compactification on an $n$-torus $T^n$), all structure constants $f^c_{ab}$ vanish,  the gauge covariant derivatives in (\ref{hyper-lagr}) become partial derivatives, and the Ricci scalar $\tilde R$  vanishes. 
  
  We use the diagonal internal metric  ansatz  \cite{CvKK}, \cite{4}:
\begin{equation}\label{gauge}
\left(\rho_ {ab} \right) = \diag \left (\rho_1, \ldots, \rho_ {n-1},
\prod_ {h = 1} ^ {n-1}
\rho_h ^ {-1} \right),
\end{equation}
and the static cylindrically  symmetric ansatzes
 for  the 4-d space-time metric and for the gauge and scalar fields
associated with the internal metric.  
We choose the  space-time metric $g_{\mu\nu}$ of the  form \eqref{cm0}.
The ansatzes for electric and magnetic fields, compatible with cylindrical symmetry, are obtained  by using the Yang--Mills equations
$\nabla_ \mu (e^{\alpha\sigma}\rho_a F ^{a\  \mu \nu})= 0$ derived from the Lagrangian \eqref{hyper-lagr}
and are of the following three forms:
$$F^a_{tu}=\frac{Q^a_e e^{2\gamma(u)}}{e^{\alpha\sigma(u)}\rho_a(u)}\equiv \tilde E^a(u)e^{2\gamma(u)}, \quad F^a_{z\phi }=Q^a_m\quad \mbox{
for electric and magnetic radial fields,}$$
$$F^a_{ tz}=Q^a_e,  \quad  F^a_{u\phi}=\frac{Q^a_m e^{2\beta(u)}}{e^{\alpha\sigma(u)}\rho_a(u)}\equiv \tilde H^a(u)e^{2\beta(u)}
\quad \mbox{for electric and magnetic longitudinal fields},$$
$$F^a_{ t\phi}=Q^a_e,  \quad  F^a_{uz}=\frac{Q^a_m e^{2\xi (u)}}{e^{\alpha\sigma(u)}\rho_a(u)}\equiv \tilde B^a(u)e^{2\xi (u)}
\quad \mbox{for electric and magnetic azimuthal fields},$$
where  $a=4,\ldots, n+3$, the constant $Q^a_m$ is the  magnetic charge and the constant $Q^a_e$  is  the  electric charge  of the configuration,
  the other components of the strength fields $F^a_{\mu\nu} $ are  equal to zero (see arguments  after \eqref{el1} and \eqref{long}).

  The allowed charge configurations are restricted by the special choice of internal metric \eqref{gauge}. Really, the Euler--Lagrange equations for  $\rho_{ab}(u)$, derived  from the Lagrangian \eqref{hyper-lagr}, read:
$$ e^{\alpha\sigma(u)+2\gamma(u)}[Q^a_mQ^b_m-\tilde E^a(u)\tilde E^b(u)]+\lambda \rho^{ab}(u)=\frac{d^2}{du^2}\rho^{ab}(u) \quad \mbox{
for the radial fields},$$
$$ e^{\alpha\sigma(u)+2\beta(u)}[\tilde H^a(u)\tilde H^b(u)-Q^a_eQ^b_e]+\lambda \rho^{ab}(u)=\frac{d^2}{du^2}\rho^{ab}(u) \quad \mbox{
for the longitudinal  fields},$$
$$ e^{\alpha\sigma(u)+2\xi (u)}[\tilde B^a(u)\tilde B^b(u)-Q^a_eQ^b_e]+\lambda \rho^{ab}(u)=\frac{d^2}{du^2}\rho^{ab}(u) \quad \mbox{
for the azimuthal  fields}.$$
  It follows from  
this that for  the diagonal metric \eqref{gauge}  the following constraints have to be satisfied:
 $$ Q^a_eQ^b_e-e^{2\alpha\sigma}\rho_a\rho_bQ^a_mQ^b_m=0  \quad \mbox{
for the radial  fields}, \quad \mbox{when} \  a\neq b,$$
$$ Q^a_mQ^b_m-e^{2\alpha\sigma}\rho_a\rho_bQ^a_eQ^b_e=0  \quad \mbox{
for the longitudinal  and azimuthal fields}, \quad \mbox{when} \  a\neq b,$$
whence for radial fields (i)\ $ Q^a_eQ^b_e=Q^a_mQ^b_m=0$ if $a\neq b$, or, generally, (ii)\ $Q^a_eQ^b_e-\kappa_a\kappa_b Q^a_mQ^b_m=0$ if $a\neq b$,  with $\kappa_a \equiv e^{\alpha\sigma}\rho_a=\const$. The latter case
  would imply the equation of motion for $e^{\alpha\sigma}\rho_a$ with $Q^a_e=Q^a_m= 0$.
  Thus,  the constraint (ii) reduces to the subset of constraints (i) which imply that the same (electric or magnetic) type of charge can appear in at most one gauge field.  The same result is valid for longitudinal and azimuthal electric and magnetic fields. Consequently, the internal isometry group $U(1)^n$ is broken down to at most $U (1) \times U (1)$, with only one electric and one magnetic charge. 
Without loss of generality,  we associate two $U(1)$ factors of the internal isometry group $U(1) \times U(1)$  with  the $(n -1)$-th and the $n$-th internal dimensions, i.e., with gauge fields $A^{n-1}_{\mu}$ and $A^{n}_{\mu}$. When
 the first $(n - 2)$ gauge fields are turned off the first $(n - 2)$ components of the diagonal internal metric become constant: $e^{2\sigma/(n\alpha)}\rho_a= \const$,   $a = 1, \ldots n - 2$.
  As a result  the solutions of $(4+n)$-dimensional Kaluza--Klein theory  are those of effective six-dimensional Kaluza--Klein theory with action:
 \begin{equation}\label{u}
S_4=\frac{1}{16\pi} \int \sqrt {-g} [R-2 (\nabla \psi) ^ 2-4 (\nabla
\chi) ^ 2
-e ^ {2 \sqrt {2} (\psi + \chi)} K ^ {\mu \nu} K_ {\mu \nu}-e ^ {2 \sqrt {2}
(\psi-\chi)} F ^ {\mu \nu} F_ {\mu \nu}]d^4x,
\end{equation}
which is obtained from \eqref{hyper-lagr} by using  the field redefinition:
$
\chi_a\equiv (1/\sqrt{2})[\ln\rho_a+2\sigma/(n\alpha)]=\const,$ $ a=1,\ldots, n-2,$ $ \chi_{n-1}\equiv (1/\sqrt{2})[\ln\rho_{n-1}+(2-n)\sigma/(n\alpha)]\equiv \chi,$ $\chi_n=-\chi,$ $ \psi \equiv \sqrt{2}\alpha\sigma, $ $
F^{n-1}_ {\mu \nu}\equiv 2 K_ {\mu \nu},$  $ F^n_ {\mu \nu}\equiv 2F_ {\mu \nu}$
 (cf. \cite{CvKK},  see also \cite{5}, \cite{4} ). 
 
 Note that the above result can also be obtained by solving the Killing spinor equations for supersymmetric Kaluza--Klein configurations with cylindrical symmetry (the proof will be given in  a separate paper).

\section{Exact solutions of Einstein--Yang--Mills--dilaton equations  with radial electric and magnetic fields}

In this section we will obtain the  Einstein equations,
Yang--Mills equations for the gauge fields $ A_ \mu $, $ B_ \nu $,
and the equations for the dilaton  $ \psi $ and scalar field $ \chi $. We will  find
exact solutions of the Einstein--Yang--Mills--dilaton (EYMD) equations
in the case of the radial electric and magnetic fields.

We consider the four-dimensional action \eqref{u} of 6-d Kaluza--Klein theory with
\begin{equation} \label{FK}
 F_ {\mu \nu} =  \partial_ {\mu} A_ {\nu} - \partial_ {\nu} A_ {\mu},\quad
 K_ {\mu \nu} = \partial_ {\mu} B_ {\nu} - \partial_ {\nu} B_ {\mu},
\end{equation}
 where $ A_ \mu $, $ B_ \nu $ are the  two Abelian gauge fields,  and  
$ \psi $ (dilaton) and  $ \chi $ are two scalar fields.
The non-zero components of the Ricci tensor $ R_ \nu ^ \mu $ and the
Riccci scalar $R\equiv R_\mu^\mu$ of the metric~\eqref{cm0} are:
$$\begin{array}{l}
R_t ^ t =  e ^ {-2(\beta+ \gamma+\xi) }\gamma'',
\\ \\
R_u ^ u = e ^ {-2(\beta+ \gamma+\xi) } [\gamma'' + \xi'' + \beta'' - 2 (\beta
'\xi' +
\xi '\gamma' + \gamma '\beta')],
\\ \\
R_z ^ z = e ^ {-2(\beta+ \gamma+\xi) }\xi'' ,
\\ \\
R_\phi ^ \phi = e ^ {-2(\beta+ \gamma+\xi) }\beta'' ,\\ \\
 R=2e ^ {-2(\beta+ \gamma+\xi) }(\gamma'' + \xi'' + \beta'' -
\beta '\xi' -
\xi '\gamma' - \gamma '\beta'),
\end{array}$$
where the prime denotes differentiation with respect to $ u $.
The Euler--Lagrange equations for  the scalar fields $ \psi $ and $ \chi $, derived  from the the action~\eqref{u}, are given by:
\begin{equation} \label{dilp}\begin{array}{l}
\Box \psi = (1/\sqrt{2}) \left(e ^ {2 \sqrt {2} (\psi +
\chi)} K_ {\mu \nu} K ^ {\mu \nu} + e ^ {2 \sqrt {2} (\psi-\chi)} F_ {\mu \nu} F
^ {\mu \nu}\right),\\ \\
\Box \chi = (1/(2 \sqrt{2}))
\left(e ^ {2 \sqrt {2} (\psi +
\chi)} K_ {\mu \nu} K ^ {\mu \nu} - e ^ {2 \sqrt {2} (\psi-\chi)} F_ {\mu \nu} F
^ {\mu \nu}\right).
\end{array}\end{equation}
$\left(\Box\equiv g^{\mu\nu}\nabla_\mu\nabla_\nu\right)$. Variation in the fields $ A_ \nu $, $ B_ \nu $ gives the Yang--Mills equations:
\begin{equation} \label{YMeq}
\nabla_ \nu F ^ {\mu \nu}  = 2 \sqrt {2}  e ^ {2 \sqrt {2}(\psi-\chi)} F ^ {\mu \nu}\nabla_\nu (\chi-\psi), \quad
\nabla_ \nu K ^ {\mu \nu}  = - 2 \sqrt {2}  e ^ {2 \sqrt {2}(\psi+\chi)} K ^ {\mu \nu}\nabla_\nu (\psi+\chi),
\end{equation}
where $\nabla_ \nu$ is covariant derivative with respect to $x^\nu$.

Finally, varying the metric $ g_ {\mu \nu} $, we obtain the Einstein equations
\begin{equation} \label{ur_ein1}
R_ \nu ^ \mu = \tilde{T} _ \nu ^ \mu, \quad \tilde {T} _ \nu ^ \mu: = T_\nu ^
\mu
- (1/2) \delta_ \nu ^ \mu T \qquad (T\equiv T_ \lambda ^
\lambda)\end{equation}
with the energy-momentum tensor
\begin{multline} \label{T}
T_ {\mu \nu} = 2 \nabla_ \mu \psi \nabla_ \nu \psi-g_ {\mu \nu} (\nabla \psi) ^
2 +4 \nabla_ \mu \chi \nabla_ \nu \chi
- 2 g_ {\mu \nu} (\nabla \chi) ^ 2 + \\
e ^ {2 \sqrt {2} (\psi +
\chi)} (2K_ {\mu \tau} K^{\ \tau}_\nu-(1/2) g_ {\mu \nu} K_ {\mu \nu} K ^ {\mu \nu}) +
e ^ {2 \sqrt {2} (\psi-\chi)} (2F_ {\mu \tau} F^{\ \tau}_\nu-(1/2)
g_ {\mu \nu} F_ {\mu \nu} F ^ {\mu \nu}).
\end{multline}
We expand the gauge fields $ A_\mu $ and $ B_ \mu $ into temporal and spatial
components:
$
A_ \mu = (A_t, \vec {A}), \quad B_ \mu = (B_t, \vec{B}),
$
where
$
\vec{A} = (A_u, A_z, A_ \phi), \quad \vec{B} = (B_u, B_z, B_ \phi),
$
and consider the ansatz 
for the form of the next static Abelian gauge fields, dilaton  and scalar field:
\begin{equation} \label{anzzz}
A_ \mu = (A_t (u), 0,0,0), \quad B_ \mu = (0,0,0, \mbox{const}\cdot z), \quad \psi =\psi(u),\quad \chi=\chi(u).
\end{equation}
We introduce the three-dimensional vector fields:
\begin{equation} \label{el1}
\vec {E} _A: = - \vec {\partial} A_t + \frac {\partial \vec {A}} {\partial t},
\quad \vec {H} _A: = [\vec {\partial }, \vec {A}],
\quad\vec{E} _B: = - \vec {\partial} B_t + \frac {\partial \vec {B}} {\partial t},
\quad \vec {H} _B: =
[\vec {\partial}, \vec {B}],
\end{equation}
where $ \vec {\partial} = (\partial_u, \partial_z, \partial_ \phi) $, \ $ ([\vec
{\partial}, \vec {A}]) ^ h: = e ^ {- [\gamma+2 (\xi+ \beta)]} \varepsilon ^ {hkl}
\partial_k A_l $ with  totally antisymmetric $\varepsilon ^ {hkl}$:  $\varepsilon ^ {u z \phi}=1$, and $h,k,l$ running over $
u, z, \phi $; components are defined similarly for $ [\vec {\partial}, \vec {B}]
$. By an analogy with the electromagnetic field, $ \vec {E} _A $, $ \vec {E} _B
$ are called electric fields and $ \vec {H} _A $, $ \vec {H} _B $ are called
magnetic fields.
Since from~\eqref {anzzz}--\eqref{el1} only the components $ E^u_ {A} $, $ H_B ^
u $ are non-zero, we call the fields $ A_\mu$, $ B_\mu $ radial.

Using~\eqref{FK} and \eqref{anzzz} we find the
solutions of the Yang--Mills equations~\eqref{YMeq}, describing the radial
electric and magnetic fields:
$$\begin{array}{c}
  K^ {z \phi} = Q_me^{-2(\beta+\xi)}, \quad K ^ {t \phi} = K ^ {tz} = K ^ {tu} = K ^ {uz} = K ^
{u \phi} = 0, \quad
  Q_m = \const;\\ \\
F ^ {ut} =  {Q_e} e^{-2[\beta+\gamma+\xi+ \sqrt {2} (\psi-\chi)]} , \quad F ^ {t
\phi} = F ^ {tz} = F ^ {uz} = F ^ {u \phi} = F ^ {z \phi} = 0,
\quad Q_e = \const.
\end{array}$$
The non-zero components of the tensor $ \tilde {T} _ \mu ^ \nu$ are:
$$\begin{array}{c}
 \tilde {T} _ \phi ^ \phi=\tilde {T} _ z ^ z= -\tilde{T}^t_t =
Q_e ^ 2 e ^ {-2 [\beta+\xi+ \sqrt {2} (\psi-\chi)]}+Q_m ^ 2 e ^ {-2[
\beta+\xi- \sqrt {2} (\psi + \chi)]} ,
\\ \\
 \tilde {T} _ u ^ u = -2   e ^ {-2 (\gamma+ \xi+ \beta)}\left(\psi '^ 2+2\chi'^2\right) -
Q_e ^ 2 e ^ {-2[ \beta+\xi+ \sqrt {2} (\psi-\chi)]}- Q_m ^ 2 e ^ {-2[ \beta+\xi- \sqrt {2} (\psi +
\chi)]} .
\end{array}$$
Equations~\eqref{dilp} give:
\begin{equation} \label{beta2}
 \chi'' =-(1/\sqrt {2})\left(
Q_e ^ 2 e ^ {2 [\gamma - \sqrt {2} (\psi-\chi)]}+Q_m ^ 2 e ^ {2 [\gamma + \sqrt {2} (\psi + \chi)]}\right) ,
\end{equation}
\begin {equation} \label{psi2}
  \psi'' =\sqrt {2}\left( Q_e ^ 2 e ^ {2 [\gamma-
\sqrt {2}
(\psi-\chi)]} -
Q_m ^ 2 e ^ {2 [\gamma+\sqrt {2} (\psi + \chi)]}\right).
\end{equation}
The Einstein equations $R_ \nu ^ \mu = \tilde{T} _ \nu ^ \mu$~\eqref{ur_ein1}
reduce to the next four equations:
\begin{equation}\label{gamma@@}
 \gamma'' = -
Q_e ^ 2 e ^ {2 [\gamma - \sqrt {2} (\psi-\chi)]}-Q_m ^ 2 e ^ {2 [\gamma+ \sqrt {2} (\psi + \chi)]}  \qquad (R_ t ^ t =
\tilde{T} _ t ^ t),
\end{equation}
\begin{equation}\label{xi@@}
 \xi'' =Q_e ^ 2 e ^ {2 [\gamma - \sqrt {2} (\psi-\chi)]}+Q_m ^ 2 e ^ {2 [\gamma+ \sqrt {2} (\psi + \chi)]} \qquad (R_ z ^ z =
\tilde{T} _ z ^ z),
\end{equation}
\begin{equation}\label{BETA222}
 \beta'' =Q_e ^ 2 e ^ {2 [\gamma - \sqrt {2} (\psi-\chi)]}+Q_m ^ 2 e ^ {2 [\gamma+ \sqrt {2} (\psi + \chi)]} \qquad (R_ \phi ^ \phi =
\tilde{T} _ \phi ^ \phi),
\end{equation}
\begin{equation} \label{vsp20}
\gamma'' + \xi'' + \beta'' - 2 (\beta
'\xi' +
\xi '\gamma' + \gamma '\beta') =-2 \psi '^ 2  -4\chi'^2
- Q_e ^ 2 e ^ {2 [\gamma - \sqrt {2} (\psi-\chi)]}-Q_m ^ 2 e ^ {2 [\gamma+ \sqrt {2} (\psi + \chi)]} \qquad (R^u_u=
\tilde{T}^u_u).
\end{equation}
The other six Einstein equations are satisfied identically.
Substituting~\eqref{gamma@@}--\eqref{BETA222} into~\eqref{vsp20}, we obtain
\begin{equation}\label{vsp2}
 \gamma ' \beta' + \xi' \gamma' + \beta ' \xi' = \psi '^ 2 + 2\chi'^2 +Q_e ^ 2 e ^ {2 [\gamma - \sqrt {2} (\psi-\chi)]}+ Q_m ^ 2
e ^ {2 [\gamma+ \sqrt{2} (\psi + \chi)]}
 .
\end{equation}
 We notice that the
equation obtained by differentiating 
~\eqref{vsp2} is satisfied
identically as consequence of~\eqref{beta2}--\eqref{BETA222}, 
which imposes
restrictions on the initial data for 
this system.
Integrating equations~\eqref{beta2}, \eqref{gamma@@} we see that  $ \gamma'' =
\sqrt {2} \chi'' $. Setting the constants of
integration equal to zero, we obtain
\begin {equation} \label{rad1}
\gamma = \sqrt {2} \chi.
\end{equation}
In addition, from~\eqref{gamma@@}--\eqref{BETA222} we get:
$\beta = - \gamma + a u + \beta_0, \quad \xi = - \gamma + bu + \xi_0$, where $ a $, $ d $, $ \beta_0 $, $ \xi_0 $ are the constants of integration. By scale transformations of coordinates $z$ and $\phi$ the additive constants $ \beta_0 $, $ \xi_0 $ can be made to vanish, so we have

\begin{equation} \label{rad2}
\beta = - \gamma + a u, \quad \xi = - \gamma + bu .
\end{equation}

From~\eqref{psi2}, \eqref{gamma@@} and \eqref{rad1} it follows that
\begin {equation} \label{nov1}
   \gamma'' =-Q_e ^ 2 e ^ {2(2 \gamma- \sqrt {2} \psi)}- Q_m ^ 2 e ^ {2(2 \gamma+ \sqrt {2} \psi)}
, \quad
   \psi'' =\sqrt {2} \left(Q_e ^ 2 e ^ {2(2\gamma- \sqrt {2} \psi)}
-
Q_m ^ 2 e ^ {2(2 \gamma+ \sqrt {2} \psi)}\right),
\end{equation}
supplemented by the condition~\eqref{vsp2}, which in view of~\eqref{rad1}
and~\eqref{rad2} takes the form
\begin {equation}
-2 \gamma'^ 2 + ab = \psi'^ 2 + Q_m ^ 2 e ^ {2(2 \gamma+ \sqrt {2} \psi)}
+ Q_e ^ 2 e ^ {2(2 \gamma- \sqrt {2} \psi)}.
\end{equation}
 The system \eqref{nov1} 
is equivalent to the following 
system:
\begin {equation} \label{nov_eta}
 \eta'' =-2Q_e ^ 2 e ^ {4\eta}, \quad  \zeta'' =-2Q_m ^ 2 e ^ {4\zeta},
\end{equation}
where
\begin {equation} \label{gam_psi}
   \eta =\gamma-(1/\sqrt {2})\psi, \quad \zeta =\gamma+(1/\sqrt {2})\psi.
\end{equation}

Consider all possible cases: 1) $ Q_e Q_m \neq 0 $, 2)
 $ Q_e \neq 0 $, $ Q_m = 0 $,
 3) $ Q_e = 0 $, $ Q_m
\neq 0 $, and 4) $ Q_e = Q_m = 0 $.

 Solving the system~\eqref{nov_eta} in the  case $ Q_e Q_m \neq 0 $  and taking into account that, by \eqref{gam_psi}, $\gamma=(\zeta+\eta)/2$, $\psi=(\zeta-\eta)/\sqrt{2}$, we find
 \begin{equation} \label{gamma_r1}
 \gamma (u) = - (1/4) \ln \left (4\left |q_e q_m \right |\cosh [h_e (u-u_e)] \cosh [h_m (u-u_m)]  \right),
 \end {equation}
 \begin{equation} \label{psi_r1}
 \psi (u) = -(1/( 2 \sqrt {2})) \ln \left (\left |q_m/q_e \right | \cosh [h_m
(u-u_m)]/\cosh [h_e (u- u_e)] \right),
 \end{equation}
where $ h_e\neq 0 $, $ h_m\neq 0 $, $ u_e $, $ u_m $ are  constants of integration, $Q_e/h_e\equiv q_e  $ and $Q_m/h_m\equiv q_m  $. Without
loss of generality we can assume that  $ h_e>0 , h_m >0$.
 From~\eqref{rad2} we obtain:
 \begin{equation} \label{beta_r1}
\beta (u) = (1/4) \ln \left (4\left |q_e q_m\right |\cosh [h_e (u-u_e)] \cosh [h_m (u-u_m)]  \right) + au,
\end{equation}
\begin{equation} \label {xi_r1}
 \xi (u) = (1/4) \ln \left (4\left | q_e q_m \right |\cosh [h_e (u-u_e)] \cosh [h_m (u-u_m)]  \right) + bu.
 \end{equation}
 Substituting \eqref{gamma_r1}--\eqref{xi_r1} into~\eqref {vsp2},
we have
$$ 
 4ab = h_e ^ 2 + h_m ^ 2.
 $$
From~\eqref{beta_r1} we derive
\begin{equation} \label {diffbeta}
\beta'(u) =(1/4) \left (h_e \tanh [h_e (u-u_e)] + h_m \tanh [h_m (u-u_m)]
\right) + a.
\end{equation}
Due to the monotonicity of $ \tanh $ the derivative 
$\beta'(u) $
can vanish at no more than one point.
If
\begin{equation} \label{uslgorl}
| a | < (h_e + h_m)/4
\end{equation}
then there exists a (unique) point $ u_0 \in \mathbb {R} $ for which $
\beta'(u_0) = 0 $.
If the condition~\eqref{uslgorl} is not satisfied, then there is no point at
which the derivative
\eqref{diffbeta} vanishes.

 Note that because of the transcendence of the equation $ \beta'(u) = 0 $ the
value $ u_0 $ can be found analytically only for some particular  values  of the parameters 
$h_e$, $h_m$.

Since, by~\eqref{BETA222}, the second derivative of $ \beta (u) $ is positive,
taking into account the condition~\eqref {uslgorl}, the function $ \beta(u) $
has an 
absolute minimum at $ u = u_0 $. In accordance with 
definition~\ref {opr1}, metric~\eqref{cm0}
with functions~\eqref{gamma_r1}, \eqref {beta_r1}, \eqref {xi_r1} together with
condition~\eqref{uslgorl} describes a 
family of cylindrically
symmetric wormholes characterized by an
electric charge $ Q_e $, a magnetic charge $ Q_m $ and parameters $ h_e $, $ h_m
$, $ a $ and $ b $. Note that this solution is not asymptotically flat.

We put $r_e=\exp{(h_eu_e)}$, $r_m=\exp{(h_mu_m)}$ and
introduce a new ``radial'' coordinate $ r: = \exp{(u)} \in [0, + \infty) $.
With these 
new coordinates the metric~\eqref{cm0} with the functions
\eqref{gamma_r1}, \eqref{beta_r1}, \eqref{xi_r1} takes the form
 \begin{equation}\label{novcoord}
 ds^2 = \kappa \Omega \left( \frac{dt^2}{\kappa^2\Omega^2}
 - r^{2(a+b-1)} d r^2 -  r^{2b} d {z}^2  -  r^{2a}  d {\phi}^2\right),
 \end{equation}
where
$$\kappa=\sqrt{|q_eq_m|}, \quad
\Omega=\sqrt{[(r/r_e)^{h_e} +
(r/r_e)^{-h_e}][(r/r_m)^{h_m} +
(r/r_m)^{-h_m}]},\quad 4ab=h_e^2+h_m^2$$
and $
h_eh_mQ_eQ_m\neq 0$. If \  $4| a | <h_e + h_m$ then \eqref{novcoord} is the metric of the wormhole with the throat radius $r_0$   defined by the equation
$$
(4a+h_e+h_m)({r_0}/{r_e})^{2h_e}\left({r_0}/{r_m}\right)^{2h_m} +(4a-h_e+h_m)({r_0}/{r_m})^{2h_m}+(4a+h_e-h_m)({r_0}/{r_e})^{2h_e}+
1=0.
$$
The wormhole \eqref{novcoord} with $4| a | <h_e + h_m$ which we  denote by  $\rm{WhCR^{e;m}}$ is generated by the following  
Abelian gauge fields and scalar fields:
\begin{equation}\label{A_mu_1}\begin{array}{c}
A_ \mu = \left(-(h_e/(4q_e))\left[(r/r_e)^{h_e} -
(r/r_e)^{-h_e}\right]/ |\left[(r/r_e)^{h_e} +
(r/r_e)^{-h_e}\right],\ 0,\ 0,\ 0\right), \quad
B_ \mu = (0,\ 0,\ 0,\ -h_mq_m z),\\ \\
F^{\mu \nu}=(h_e/q_e)\left[(r/r_e)^{h_e} +
(r/r_e)^{-h_e}\right]^{-2}\delta^\mu_u\delta^\nu_t, \quad
K_{\mu \nu}=h_mq_m\delta_\mu^z\delta_\nu^t,\\ \\
 \psi =(1/(2\sqrt{2}))\ln\left(\left|q_e/q_m\right|[(r/r_e)^{h_e} +
(r/r_e)^{-h_e}]/[(r/r_m)^{h_m}+(r/r_m)^{-h_m}]\right),\\ \\
\chi=-(1/(4\sqrt{2}))\ln\left(\left|q_eq_m\right|[(r/r_e)^{h_e} +
(r/r_e)^{-h_e}][(r/r_m)^{h_m}+(r/r_m)^{-h_m}]\right).
\end{array}\end{equation}

 In the second case when $ Q_e \neq 0$ and $ Q_m = 0 $
we obtain:
$$
\gamma = \sqrt{2}\chi= - (1/4) \ln D(u)
 + c u,\ \beta = -\gamma +  a u, \  \xi =   -\gamma +  b u, \ \psi  =  (1/(2\sqrt{2}))\ln D(u) +\sqrt{2}\ c u ,
$$
where
$D(u)\equiv 2\left |q_e \right |\cosh [h_e (u-u_e)]
$
and, by \eqref{vsp2},
$
4(ab-4c^2) = h_e^2.
$
We have
$$
\beta' (u) = (h_e/4)  \tanh[h_e( u-u_e)]  + a - c.
$$
The single point at which the derivative of $ \beta (u)$ vanishes, exists only if \ $
4| a-c | < h_e $, and it is given by
$$
u_0 = u_e - (1/h_e) \artanh[4(a-c
)/h_e].
$$
The hypersurface $ u = u_0 $ defines a
throat of the cylindrically symmetric wormhole, which we denote by $\rm WhCR^e$.
Its metric in coordinates $t,r,z,\phi$ is
 \begin{equation}\label{CWH_2}
 ds^2 =  k_e\Lambda_e \left( \frac{r^{2c}dt^2}{k_e^2\Lambda_e^2}
 - r^{2(a+b-c-1)} d r^2 -  r^{2(b-c)} d {z}^2  -  r^{2(a-c)}
  d {\phi}^2\right)\quad (a,b,c 
= \mathrm{const}),
 \end{equation}
 where
\begin{equation}\label{LamR2}
k_e=\sqrt{|q_e|}, \quad
\Lambda_e=\sqrt{(r/r_e)^{h_e} +
(r/r_e)^{-h_e}},\quad 4ab=h_e^2+16c^2 \quad
(h_e, q_e  = \mathrm{const},h_eq_e\neq 0).
\end{equation}
  The throat radius  of the $\rm WhCR^e $ is
  \begin{equation}\label{ThrR2}
  r_0=r_e\left(\frac{h_e-4(a-c)}{h_e+4(a-c)}\right)^{1/(2h_e)}, \quad h_e>4|a-c|,
\end{equation}
and the wormhole
 is generated by  the following single   Abelian gauge field $A_\mu$,
 dilaton  $\psi$ and scalar field  $\chi$:
\begin{equation}\label{dilR2}\begin{array}{c}
A_ \mu = \left(-(h_e/(4q_e))\left[(r/r_e)^{h_e} -
(r/r_e)^{-h_e}\right]\left[(r/r_e)^{h_e} +
(r/r_e)^{-h_e}\right]^{-1},\ 0,\ 0,\ 0\right), \\ \\
F^{\mu \nu}=(h_e/q_e)\left[(r/r_e)^{h_e} +
(r/r_e)^{-h_e}\right]^{-2}\delta^\mu_u\delta^\nu_t, \\ \\
 \psi =(1/(2\sqrt{2}))\ln\left(|q_e|
 r^{4c}[(r/r_e)^{h_e} +
(r/r_e)^{-h_e}]\right),\ \chi=-(1/(4\sqrt{2}))\ln\left(|q_e|
 r^{-4c}[(r/r_e)^{h_e} +
(r/r_e)^{-h_e}]\right).\end{array}\end{equation}

The wormhole $\rm WhCR^m$ corresponding to the third case ($Q_e=0$,  $Q_m\neq 0$) is defined by the metric 
\begin{equation}\label{CWH_2m}
 ds^2 =  k_m\Lambda_m \left( \frac{r^{2c}dt^2}{k_m^2\Lambda_m^2}
 - r^{2(a+b-c-1)} d r^2 -  r^{2(b-c)} d {z}^2  -  r^{2(a-c)}
  d {\phi}^2\right)\quad (a,b,c = \mathrm{const}),
 \end{equation} where
\begin{equation}\label{LamR3}
k_m=\sqrt{|q_m|}, \
\Lambda_m=\sqrt{(r/r_m)^{h_m} +
(r/r_m)^{-h_m}},\ 4ab=h_m^2+16c^2 \quad(h_m, q_m = \mathrm{const},h_mq_m\neq 0)
, \end{equation}
 and by the formula
\begin{equation}\label{dilR}\begin{array}{c}
B_ \mu = (0,\ 0,\ 0,\ -h_mq_m z),\quad
K_{\mu \nu}=h_mq_m\delta_\mu^z\delta_\nu^t,\\ \\
 \psi =-(1/(2\sqrt{2}))\ln\left(|q_m|r^{4c}
[(r/r_m)^{h_m} +
(r/r_m)^{-h_m}]\right),\\ \\
 \chi=-(1/(4\sqrt{2}))\ln\left(|q_m|r^{-4c}
 [(r/r_m)^{h_m} +
(r/r_m)^{-h_m}]\right), \\ \\
  r_0=r_m\left(\frac{\displaystyle h_m-4(a-c)}{\displaystyle h_m+4(a-c)}\right)^{1/(2h_m)}, \quad h_m>4|a-c|
.  \end{array}\end{equation}

Finally, in the  case $ Q_e = Q_m = 0 $ we have non-wormhole exact  solutions of the Einstein-dilaton-scalar field equations:
 \begin{equation}\label{non_WH}\begin{array}{c}
 ds^2 =  r^{2c}dt^2
 - r^{2(a+b+c-1)} d r^2 -  r^{2b} d {z}^2  -  r^{2a}
  d {\phi}^2\quad (a,b,c = \mathrm{const},\  r\in (0,+\infty)),\\ \\
\psi=k_1\ln r+k_0, \quad \chi=s_1\ln r+s_0, \quad ab+ac+bc=k_1^2+2s_1^2\quad
(k_0,k_1,s_0,s_1 = \mathrm{const}).
 \end{array}\end{equation}

Let's consider the question of the stability of the solutions.
 Introducing the variables $ \gamma_1: = \gamma'$, $ \psi_1: = \psi' $, we
rewrite the system~\eqref{nov1} in the form:
 \begin{equation}\label{zeta1}
  \left\{
 \begin{aligned}
 \gamma' &=\gamma_1,\\
 \gamma_1' &=
Q_e ^ 2 e ^ {2(2 \gamma- \sqrt {2} \psi)}+Q_m ^ 2 e ^ {2 (2\gamma+ \sqrt {2} \psi)} , \\
 \psi' &=\psi_1,\\
 \psi_1' &=\sqrt{2} \left(Q_e ^ 2 e ^ {2(2 \gamma - \sqrt {2} \psi)} -
Q_m ^ 2 e ^ {2( 2\gamma + \sqrt {2} \psi)}\right).
 \end{aligned}
 \right.
 \end{equation}
 Given that the equilibrium point  of 
a system:
$ d\vec{x}/d\tau = \vec{f}(\vec{x}(\tau)),
 $
\ where $\vec{x}$, $\vec{f}$ are vector strings, is the point at which the
right-hand side vanishes: $ \vec{f}(\vec{x})=0$ \cite{ms}, we find that the equilibrium points of
the system \eqref{zeta1} are defined by the  equations:
$
  \gamma_1=\psi_1=
 Q_e^2 e ^ {2(2 \gamma- \sqrt {2} \psi)}= \ Q_m^2 e ^ {2(2 \gamma+
\sqrt {2} \psi)}=0.
 $
We can see from here that, if at least one of the charges $ Q_e $ or $ Q_m $ is nonzero,
then the equilibrium points are absent, indicating  the stability of our
wormhole solutions.

In this section we have found three new types of static cylindrically symmetric  dilatonic wormholes: dyonic  ${\rm WhCR^{e;m}}$ defined by   \eqref{novcoord} -- \eqref {A_mu_1}, 
$\rm WhCR^e$  \eqref{CWH_2} -- \eqref{dilR2} with nonzero electric charge \
and  $\rm WhCR^m$  \eqref{CWH_2m} -- \eqref{dilR} with nonzero magnetic charge.
All found solutions are Jacobi stable. 

The purely dilatonic exact solutions  determined by \eqref{non_WH}  turn out to be non-wormholes.
So   the wormhole solutions  are not available when both electric and magnetic charges are equal to
zero. Hence, one can conclude
that the  wormhole properties of our 
solutions are generated by electric and/or magnetic
charges of static Abelian gauge fields.

\section {Exact solutions of Einstein--Yang--Mills--dilaton equations with longitudinal electric and magnetic fields}
In this section we write and integrate the  Einstein--Yang--Mills equations
and dilaton and scalar field equations for the following ansatz:
\begin{equation}\label{long}
A_\mu = (\const \cdot z,0,0,0), \quad B_\mu = (0,0,0,B_\phi (u)), \quad \psi=\psi(u), \quad \chi=\chi(u).
\end{equation}
Of the components~\eqref{el1} only  $ E^z_ {A} $ and $ H_B ^ z $ are 
non-zero, so one can speak of 
longitudinal electric and magnetic fields.
Using \eqref{FK} and solving  the Yang--Mills equations~\eqref{YMeq} we have
$$
  K^{u\phi}=Q_m e^{-2[\beta+\gamma+\xi +\sqrt{2}(\psi+\chi)]}, \quad
K^{u\phi}=K^{tz}=K^{tu}=K^{uz}=K^{u\phi}=0
  , \quad
  Q_m=\const;
$$
$$ F^{zt}=Q_ee^{-2(\gamma+\xi)}, \quad F^{t\phi}=F^{uz}=F^{ut}=F^{u\phi}=F^{z\phi}=0,
 \quad Q_e=\const.
$$
The equations~\eqref{dilp} give:
\begin{equation} \label{beta3}
 \chi'' =-(1/\sqrt {2})\left(Q_e^2  e^{2[\beta+\sqrt{2}(\psi-\chi)]}+Q_m^2
e^{2[\beta-\sqrt{2}(\psi+\chi)]}
\right),
\end{equation}
\begin {equation} \label{psi3}
  \psi'' =\sqrt {2}\left(Q_e^2  e^{2[\beta+\sqrt{2}(\psi-\chi)]}-Q_m^2
e^{2[\beta-\sqrt{2}(\psi+\chi)]}
\right).
\end{equation}
The non-trivial Einstein equations~\eqref{ur_ein1} are

\begin{equation}\label{gamma@@3}
 -\gamma'' = Q_e^2  e^{2[\beta+\sqrt{2}(\psi-\chi)]}+Q_m^2
e^{2[\beta-\sqrt{2}(\psi+\chi)]}
,
\end{equation}
\begin{equation}\label{xi@@3}
 -\xi'' = Q_e^2  e^{2[\beta+\sqrt{2}(\psi-\chi)]}+Q_m^2
e^{2[\beta-\sqrt{2}(\psi+\chi)]},
\end{equation}
\begin{equation}\label{BETA2223}
 \beta'' = Q_e^2  e^{2[\beta+\sqrt{2}(\psi-\chi)]}+Q_m^2
e^{2[\beta-\sqrt{2}(\psi+\chi)]},
\end{equation}
\begin{equation}\label{vsp30}
 \gamma'' + \xi'' + \beta'' - 2 (\beta
'\xi' +
\xi '\gamma' + \gamma '\beta') =-2 \psi '^ 2  - 4\chi'^2 +
Q_m^2e^{2\beta}e^{-2\sqrt{2}(\psi+\chi)} +
Q_e^2  e^{2\beta}e^{2\sqrt{2}(\psi-\chi)}.
\end{equation}
From~\eqref{gamma@@3}--\eqref{vsp30} we get
\begin{equation} \label{vsp3}
 \gamma' \beta' + \xi' \gamma' + \beta' \xi' = \psi'^2 + 2\chi'^2 - Q_m^2
e^{2\beta} e^{-2\sqrt{2}(\psi+\chi)}
- Q_e^2  e^{2\beta}e^{2\sqrt{2}(\psi-\chi)}.
\end{equation}
By virtue of~\eqref{beta3}--\eqref{BETA2223} the differential consequence of the last equation is satisfied
identically, 
which imposes
restrictions on the initial
data for the system.
From~\eqref{beta3}, \eqref{BETA2223} it follows  $
\beta'' = -\sqrt {2}
\chi'' $. We put
\begin {equation} \label{rad12}
\beta = -\sqrt {2} \chi
\end{equation}
and, from~\eqref{gamma@@3}--\eqref{BETA2223},
\begin{equation}\label{rad22}
\gamma = -\beta + a u , \quad \xi = - \beta + b u \qquad (a,b = \const),
\end{equation}
where the additive integration constants  are eliminated by rescaling of  $t,\ z$.
The system~\eqref{beta3}--\eqref{BETA2223} reduces to the form:
\begin {equation} \label{beta_nov1}
   \beta'' =Q_e ^ 2 e ^ {2(2 \beta- \sqrt {2} \psi)}+ Q_m ^ 2 e ^ {2(2 \beta- \sqrt {2} \psi)}
,\quad
    (1/\sqrt {2})\psi'' = Q_e ^ 2 e ^ {2(2 \beta- \sqrt {2} \psi)}- Q_m ^ 2 e ^ {2(2 \beta- \sqrt {2} \psi)}.
\end{equation}
 From~\eqref{vsp3} we have
 \begin{equation}\label{vsp4}
 Q_e^2  e^{2(2\beta+\sqrt{2}\psi)}+ Q_m^2 e^{2(2\beta-\sqrt{2}\psi)} + ab = \psi'^2 +2\beta'^2
.
\end{equation}
In variables $\bar\eta=\beta+(1/\sqrt{2})\psi$, $\bar\zeta=\beta-(1/\sqrt{2})\psi$  \eqref{beta_nov1} takes the form
 \begin{equation}\label{bar_eta} 
 \bar\eta ''=2Q_e^2e^{4\bar\eta}, \quad \bar\zeta ''=2Q_m^2e^{4\bar\zeta}.
 \end{equation}

Assume 
$ Q_e Q_m \neq 0 $.
 From \eqref{bar_eta} it follows
 $$ \bar\eta '{}^2-Q_e^2e^{4\bar\eta}=(\varepsilon/4) h_e^2, \quad \bar\zeta '{}^2-Q_m^2e^{4\bar\zeta}=(\bar\varepsilon/4) h_m^2,$$
 where $h_e,\ h_m$ are positive integration constants and $\varepsilon, \bar\varepsilon $
take values $0, \pm 1$.
  Considering all possible cases we obtain the 
solutions: $\bar\eta_k=-\ln \Lambda_{k|e},$ $\bar\zeta_ k=-\ln \Lambda_{k|m}$ where  $k =1,2,3$ and
    \begin{equation}\label{Lam_e}\begin{array}{c}
      \Lambda_{1|e}=\sqrt{2}\sqrt{\left|Q_e(u-u_e)\right|}, \quad
      \Lambda_{ 2|e}=\sqrt{2}\sqrt{\left|q_e\sinh[h_e(u-u_e)]\right|},\quad
      \Lambda_{ 3|e}=\sqrt{2}\sqrt{\left|q_e\sin[h_e(u-u_e)]\right|},\\ \\
  \Lambda_{1|m}=\sqrt{2}\sqrt{\left|Q_m(u-u_m)\right|}, \quad
      \Lambda_{2|m}=\sqrt{2}\sqrt{\left|q_m\sinh[h_m(u-u_m)]\right|},\quad
      \Lambda_{3|m}=\sqrt{2}\sqrt{\left|q_m\sin[h_m(u-u_m)]\right|},
   \end{array}\end{equation}
($ q_e\equiv Q_e/h_e,\ q_m\equiv Q_m/h_m,\  h_e>0,\ h_m>0$).
 Substituting these 
solutions in formulas $\beta=(\bar\eta +\bar\zeta)/2$, $\psi=(\bar\eta -\bar\zeta)/\sqrt{2}$ \ and using \eqref{rad12}, \eqref{rad22}, we obtain in the case  $ Q_e Q_m \neq 0 $ nine types 
 $(k|e;j|m) $ 
 of exact solutions of the EYMD equations marked with two indices $ k,j$ running over ${ 1,2,3}$:
   \begin{equation}\label{LonEM1}\begin{array}{c}
      ds^2_{{k|e;j|m}}=\Lambda_{ k|e}\Lambda_{ j|m}\left(e^{2au}dt^2-e^{2(a+b)u}du^2 -e^{2bu}dz^2\right)-\frac{\displaystyle d\phi^2}{\displaystyle \Lambda_{ k|e}\Lambda_{ j|m}} ,\quad 4ab=\varepsilon_kh_e^2+\bar \varepsilon_j h_m^2,  \\ \\
   \psi_{k|e;j|m}=(1/\sqrt{2})\ln\left(\Lambda_{ j|m}/\Lambda_{ k|e}\right),\quad
      \chi_{k|e;j|m}=\left(1/(2\sqrt{2})\right)\ln\left(\Lambda_{ k|e}\Lambda_{ j|m}\right),\\ \\
  A_\mu=(-h_eq_e z,0,0,0),\quad B_\mu^{j|m}=\left(0,0,0,B_\phi^{ j|m}\right),\\ \\
\end{array}\end{equation}
here $\varepsilon_1=\bar \varepsilon_1=0,\varepsilon_2=\bar \varepsilon_2=1,\varepsilon_3=\bar \varepsilon_3=-1$ and
  \begin{equation}\label{B_phi}\begin{array}{c}
  B_\phi^{ 1|m}=-(1/(4Q_m))(u-u_m),\ B_\phi^{ 2|m}=-(1/(4q_m))\coth[h_m(u-u_m)],\\ \\
  B_\phi^{ 3|m}=-(1/(4q_m))\cot[h_m(u-u_m)].
\end{array}\end{equation}

In the same way we obtain   three  types 
$(k|e)$, $k =1,2,3$, 
of exact solutions   of the EYMD equations in the case $ Q_e
\neq 0 $, $ Q_m = 0 $:
\begin{equation}\label{LonE1}\begin{array}{c}
      ds^2_{ k|e}=e^{cu}\Lambda_{ k|e}\left(e^{2au}dt^2-e^{2(a+b)u}du^2 -e^{2bu}dz^2\right)-\frac{\displaystyle e^{-cu} d\phi^2}{\displaystyle  \Lambda_{ k|e}},\quad 4(ab-c^2)= \varepsilon_kh_e^2,
      \\ \\
   \psi_{{ k|e}}=(1/\sqrt{2})\left(cu-\ln\Lambda_{ k|e}\right),\quad
      \chi_{ k|e}=(1/(2\sqrt{2})\left(\ln\Lambda_{ k|e}+cu\right),\\ \\
  A_\mu=(-h_eq_e z,0,0,0),\quad B_\mu=0,\quad       \varepsilon_1=0,\ \varepsilon_2=1,\ \varepsilon_3=-1,
\end{array}\end{equation}
and three  types 
$(k|m)$, $k=1,2,3$, 
of exact solutions of the EYMD equations in the case $ Q_e = 0 $, $ Q_m
\neq 0 $:
\begin{equation}\label{LonM1}\begin{array}{c}
      ds^2_{k|m}=e^{cu}\Lambda_{k|m}\left(e^{2au}dt^2-e^{2(a+b)u}du^2 -e^{2bu}dz^2\right)-\frac{\displaystyle e^{-cu} d\phi^2}{\displaystyle  \Lambda_{k|m}},
\quad 4(ab-c^2)= \bar\varepsilon_kh_m^2,
       \\ \\
   \psi_{k|m}=(1/\sqrt{2})\left(\ln\Lambda_{k|m}-cu\right),\quad
      \chi_{k|m}=(1/(2\sqrt{2})\left(\ln\Lambda_{k|m}+cu\right),
  \\ \\
  A_\mu=0,\quad B_\mu=(0,0,0,B_\phi^{k|m}),\quad \bar\varepsilon_1=0,\ \bar\varepsilon_2=1,\ \bar\varepsilon_3=-1, 
\end{array}\end{equation}
 where  $\Lambda_{k|e}, \Lambda_{k|m}, B_\phi^{k|m}$ are defined by \eqref{Lam_e}  and \eqref{B_phi}, $c=\const$, $k=1,2,3$.

 In the case $Q_e=Q_m=0$ we again obtain non-wormhole exact solutions \eqref{non_WH} of the Einstein-dilaton-scalar field equations.

Arguing similarly to the previous case, it is easy to see that all the above solutions are Jacobi
stable.

We now turn to a discussion of the wormhole  properties of the exact solutions found in this section.
We require that the radicand of $\Lambda_{k|e}$, $\Lambda_{k|m}$ \eqref{Lam_e},  $k=1,2,3$,
  be nonzero. 
  The sets $ U_{k|e}$ of the roots of the equations $\Lambda_{k|e}=0$ are $ U_{1|e}=U_{2|e}= \{u_e \} $, $ U_{3|e}= \{u_e+ \pi N/h_e\ | N \in \mathbb{Z} \} $,  and the sets $ U_{k|m}$ of the roots of the equations $\Lambda_{k|m}=0$ are  $ U_{1|m}=U_{2|m}= \{u_m \} $,  $ U_{3|m}= \{u_m+ \pi N/h_m\ | N \in \mathbb{Z} \}. $
For  $ Q_e Q_m \neq 0 $ the domains of the functions $\ln[(\Lambda_{k|e}\Lambda_{j|m})^{-1/2}]=\beta(u) $ and  $ \psi_{k|e;j|m}$   \eqref{LonEM1}
are $D_{k|e ; j|m}\equiv \mathbb{R} \backslash (U_{k|e}\cup U_{j|m})$.  
Using \eqref{Lam_e} we find for the types $(j 
|e; 3 |m)$ , $j=1,2,3$:
\begin{equation}\label{der}
 \beta'(u) = \begin{cases}
-(1/4) ( (u-u_e)^{-1}+h_m \cot[h_m(u-u_m)]) \ \ \mbox{for type} \ \ {(1|e;3|m)}, \\
-(1/4) ( h_e \coth[h_e(u-u_e)]+h_m \cot[h_m(u-u_m)]) \ \  \mbox{for type} \ \ {(2|e;3|m)}, \\
-(1/4) (h_e \cot[h_e(u-u_e)]+h_m \cot[h_m(u-u_m)]) \ \ \mbox{for type} \ \ {(3|e;3|m)} \quad (h_e>0 , h_m >0).
              \end{cases}
\end{equation}
We put $V^m_N\equiv (u_m+ \pi N/h_m, u_m+ \pi (N+1)/h_m)$ for $N\in \mathbb{Z}$. 
Suppose  for $j=1,2$ we have $u_e\in V^m_{N_0}$
 for some $N_0\in \mathbb{Z}$. Then the domain $D_{j|e ; 3|m}$ of $\beta(u)$ is the set of the intervals $V_-^m \equiv (u_m+ \pi N_0/h_m, u_e)$, $V_+^m \equiv (u_e, u_m+ \pi (N_0+1)/h_m)$ and $ V_{N}^m$  \ for \ ${N\neq N_0}$: $D_{j|e ; 3|m}=  V_-^m \cup V_+^m  \cup_{N\neq N_0} V_{N}^m $.
 At the boundaries of each of the above intervals  the function $ \beta(u) $ and the  
circle radius �$ R (u) = e ^ {\beta (u)}$ tend to $+\infty$. It follows from \eqref{der} that $ \beta''(u)> 0 $
in the domain of $ \beta(u) $, therefore, $ \beta'(u)$  grows  monotonically in each interval  of $D_{j|e ; 3|m}$.
 Since the monotone continuous function $\beta'(u)$ tends to $-\infty$ at the left boundary of each interval of $D_{j|e ; 3|m}$ and tends to $+\infty$  at the right boundary of the  interval,  there exists a single point  for each interval:  $u_{0-}\in V_-^m$,  $u_{0+}\in V_+^m$, $u_{0N}\in V^m_N$, $N\neq N_0$,
 at which  $\beta'(u)$ vanishes and $ \beta(u) $ has a minimum.
According to the definitions~\ref{opr1} and~\ref{opr2} there is a throat in each
interval of $D_{j|e ; 3|m}$, and  the space--time
``consists'' of an infinite countable number  of 
``universes''~\eqref{LonEM1} with $u\in V^m_{N\in \mathbb{Z}}$,  $N\neq N_0$ or $u\in  V_-^m$ or  $u\in  V_+^m$; each  universe 
has a 
throat. Similarly, when  $u_e=u_m + \pi \tilde{N} /h_m$ for some $\tilde{N}\in \mathbb{Z}$ we obtain 
 an infinite countable number  of 
universes~\eqref{LonEM1}  with $u\in V^m_{N}$,  $N\in \mathbb{Z}$ .
 We denote these 
wormhole solutions by ${\rm WhCL}^{j|e;3|m}$,   $j = 1, 2$.  The solutions ${\rm WhCL}^{j|m;3|e}$,   $j = 1, 2$, are similarly defined.

We put   $V_N^e \equiv (u_e+ \pi N/h_e, u_e+ \pi (N+1)/h_e)$, $N\in \mathbb{Z}$.
The domain  $D_{3|e;3|m}=\cup_{N_1, N_2 \in \mathbb{Z}} (V_{N_1}^m \cap  V_{N_2}^e)$ of the function $\beta(u)$ for the solution of
type $(3|e;3|m)$ is the infinite set of 
intervals $(u^K_i,u^K_f)$, $K\in \mathbb{Z}$, where $u^K_i,u^K_f\in  U_{3|e}\cup U_{3|m}$, $u^K_i<u^K_f$ and the interval $(u^K_i,u^K_f)$ does not contain any element of $U_{3|e}\cup U_{3|m}$.
It follows from \eqref{der} that
  $ \beta(u) $ and the circle
radius �$ R (u) = e ^ {\beta (u)}$ tend to $+\infty$ at the boundaries of each  interval  $(u^K_i,u^K_f)$. As   the
monotonically increasing function
  $\beta'(u)$ tends to $-\infty$ when $u\to u^K_i+0$  and tends to $+\infty$  when $u\to u^K_f-0$, in each interval $(u^K_i,u^K_f)$  there exists a single point  $u^K_0$ 
 at which  $\beta'(u)$ vanishes and $ \beta(u) $ has a minimum. As in the previous case  the space--time
splits into multiple universes ~\eqref{LonEM1} with $u\in (u^K_i,u^K_f)$, $K\in \mathbb{Z}$; each  universe 
has a
throat defined by the equation $u=u^K_0$.
This 
wormhole solution is denoted  by ${\rm WhCL}^{3|e;3|m}$.

Consider the solutions of the  remaining types $(f|e;h|m)$, $f,h=1,2$. 
We have
 $$ \beta'(u) = \begin{cases}
-(1/4) ((u-u_e)^{-1}+ (u-u_m)^{-1}) \  \ \mbox{for type} \ \
{(1|e;1|m)},\\
-(1/4) ((u-u_e)^{-1}+ h_m \coth[h_m(u-u_m)]) \ \ \mbox{for type} \ \ {(1|e;2|m)}, \\
-(1/4) ( h_e \coth[h_e(u-u_e)]+(u-u_m)^{-1}) \ \ \mbox{for type} \ \ { (2|e;1|m)}, \\
-(1/4) (h_e \coth[h_e(u-u_e)]+ h_m \coth[h_m(u-u_m)]) \ \ \mbox{for type} \ \ {(2|e;2|m)} .
              \end{cases}
$$
 If $u_e=u_m$  then the  domain of  $\beta(u)$ is $(-\infty, \; u_e) \cup (u_e,\; +\infty)$.
As the first derivative  of the smooth function $\beta(u) $ does not vanish in the domain, it has no minimum  and the solutions of the types $(f|e;h|m)$, $f,h=1,2,$ with  $u_e=u_m$ are non-wormholes.  
In the case $u_e \neq u_m$ we have  $D_{f|e ; h|m}=(-\infty, u_i) \cup (u_i,u_f) \cup (u_f, +\infty)$ where \ $u_i=\min\{u_e,u_m\}$ and $u_f=\max\{u_e,u_m\}.$   The first derivative of the monotonically increasing function $\beta'(u)\in (-\infty, +\infty)$ vanishes at a single point $u_0 \in (u_i,u_f)$,
and the hupersurface $u=u_0$ defines a throat of the wormhole  denoted by ${\rm WhCL}^{f|e;h|m} $, $u_e \neq u_m$, $f,h=1,2$. Equation $\beta'(u) =0$  can be solved analytically  only for the wormhole ${\rm WhCL}^{1|e;1|m} $ with the throat  $u_0 = (u_e+u_m)/2$.

In the case $Q_m \neq 0$, $Q_e = 0$  the domains $D_{j|m}$ of the  function $\beta(u)$ for the solutions of the types $(j|m)$, $j=1,2,3,$ are $D_{1|m}=D_{2|m}=(-\infty, u_m) \cup (u_m,+\infty)$ and $D_{3|m}=\cup_{N \in \mathbb{Z}} V_N^m$. The first derivative of $\beta(u)$
is
\begin{equation}\label{bp}
\beta'(u) = \begin{cases}
-(1/4) ((u-u_m)^{-1}+ 2c), \quad \mbox{for type} \quad (1|m),\\
-(1/4) (h_m \coth[h_m(u-u_m)] + 2c), \quad \mbox{for type} \quad (2|m), \\
-(1/4) (h_m \cot[h_m(u-u_m)]+ 2c), \quad \mbox{for type} \quad (3|m).
             \end{cases}
\end{equation}
For 
type $(1|m)$ the throat $u=u_0\equiv u_m - (2c)^{-1}$ exists   when  \ $c>0$, $-\infty< u_0 <u_m$ \ or \  $c<0$, $u_m< u_0<+\infty$; the solution with $c=0$ is a non-wormhole.
Solutions of the type $(2|m)$ are wormholes with the throat  $u=u_0 \equiv u_m - \textrm{arcoth}(2c/h_m)$ \ only if  \ $2c/h_m>1$, $-\infty< u< u_m$ \ or \ $2c/h_m<-1$,  $u_m< u<+\infty$;  we have no wormholes when  $|2c/h_m| \leq 1$.
In the case 
$(3|m)$  we obtain an
infinite countable number  of the ``universes'' ~\eqref{LonM1} with $u\in V_{N\in \mathbb{Z}}^m$, each  contains a
throat 
$u=u_0\equiv u_m + h_m^{-1}(\pi N - \textrm{arccot}(2c/h_m))$.
We denote the above wormhole solutions by ${\rm WhCL}^{k|m}$,   $k = 1, 2, 3$.  The wormhole solutions ${\rm WhCL}^{k|e}$  \eqref{LonE1} are similarly defined.

The purely dilatonic exact solutions   \eqref{non_WH}  are non-wormholes.
So   the wormhole solutions  do not exist if both electric and magnetic charges are equal to
zero. 

Formulas~\eqref{LonEM1} in the case $ Q_e Q_m \neq 0 $,~\eqref{LonE1} in the  case $Q_e \neq 0 $, $ Q_m = 0 $ and~\eqref{LonM1} in the  case $Q_m \neq 0 $, $ Q_e = 0 $  define new cylindrically symmetric  exact solutions of
EYMD equations~\eqref{dilp}--\eqref{ur_ein1} with  
longitudinal electric and
magnetic fields  as sources; these solutions determine 
wormholes of the types ${\rm WhCL}^{j|e;k|m}$, ${\rm WhCL}^{k|e}$ and ${\rm WhCL}^{k|m}$, $j,k=1,2,3$. All of them 
  are not asymptotically flat.

\section{Conclusion}

We studied static axially (cylindrically) symmetric solutions in $4+n$-dimensional Kaluza--Klein theory with $n$ gauge fields in the case that the internal symmetry group is the maximal Abelian isometry group $U(1)^n$. It was shown that, as in the case of spherical symmetry \cite{5}, the consistency of the Euler--Lagrange equations for pure Einstein--Poincare gravity 
in $4+n$ dimensions with a
diagonal internal metric and a cylindrically symmetric ansatzes for gauge, dilaton  and internal scalar fields imposes constraints on the possible charge configurations of the solutions. Such solutions can exist only for configurations with no more than one nonzero electric $(Q_e)$ and one nonzero magnetic  $(Q_m)$  charge.  
 In fact,  $4+n$-dimensional Kaluza--Klein theory is reduced to 
an effective six-dimensional Kaluza--Klein theory \eqref{u} with two Abelian gauge fields $A_\mu,\ B_\mu$, dilaton field $\psi$ and scalar field $\chi$. 
We  solved the Einstein--Yang--Mills--Dilaton equations~\eqref {dilp}--\eqref{ur_ein1} derived from the action~\eqref{u} for the cylindrically symmetric space--time~\eqref{cm0} and found   the 
exact solutions of these equations in the
cases of (cylindrically symmetric) radial 
and longitudinal
magnetic and electric fields. 

 Following the definitions \ref{opr1} and \ref{opr2} (K.  Bronnikov and J.  Lemos ~\cite{1}) we  investigated the wormole properties of  the obtained solutions.
In the case of radial fields we  found three  types of static cylindrically symmetric  dilatonic wormholes: dyonic  ${\rm WhCR^{e;m}}$ defined by   \eqref{novcoord}--\eqref {A_mu_1}, 
$\rm WhCR^e$  \eqref{CWH_2}--\eqref{dilR2} with nonzero electric charge \
and  $\rm WhCR^m$   \eqref{CWH_2}, \eqref{LamR3}--\eqref{dilR} with nonzero magnetic charge.
 Nine  types of dyonic wormholes  
${\rm WhCL}^{k|e;j|m}$,  $k,j = 1, 2,3$, determined by \eqref{LonEM1} are found  in the case of longitudinal gauge fields as well as the wormhole  ${\rm WhCL}^{3|e}$  \eqref{LonE1} with nonzero electric charge and   the wormhole ${\rm WhCL}^{3|m}$  \eqref{LonM1} with nonzero magnetic charge.
All the 
solutions we found are Jacobi stable.

The purely dilatonic exact solutions   \eqref{non_WH}  are non-wormholes.
So    
wormhole solutions  do not exist if both electric and magnetic charges are equal to
zero. Hence,  the  wormhole properties of these 
solutions are generated only by electric and/or magnetic
charges of static Abelian gauge fields.

All obtained  wormhole solutions are  asymptotically nonflat; this confirms the "no--go" statement~\cite{1} about nonflat asymptotic behavior  of a 
 cylindrically symmetric
wormhole in the case  of everywhere nonnegative energy density of matter, i.~e. in the absence of ghost fields.

The detailed study of the structure of the  new wormhole solutions obtained in this paper will be a task for the next paper.

\subsection*{Acknowledgments}

We would like to thank   Professor
D. R. Brill  and Professor
S. V. Sushkov  for valuable  comments and for
 reading of the manuscript.

\end{document}